\pdfoutput=1
\documentclass[11pt]{article}

\usepackage{acl}

\usepackage{times}
\usepackage{latexsym}

\usepackage[T1]{fontenc}
\usepackage[utf8]{inputenc}

\usepackage{microtype}

\usepackage{inconsolata}

\usepackage{graphicx}

\usepackage{soul}
\usepackage{array}
\usepackage{multirow}
\usepackage{booktabs} % For thicker lines
\usepackage{amsmath}

\usepackage{colortbl}
\usepackage{xcolor}
\definecolor{lightgray}{rgb}{0.9, 0.9, 0.9}
\definecolor{lightred}{rgb}{1.0, 0.9, 0.9}
\usepackage{dcolumn}
\usepackage{siunitx}
\usepackage{etoolbox}
\usepackage{dblfloatfix}
\usepackage{makecell}

\title{Fine-Tuning Whisper for Inclusive Prosodic Stress Analysis}

\author{Samuel S. Sohn \\
  Department of Psychology \& \\
  Center for Cognitive Science\\
  Rutgers University --\\
  New Brunswick \\
  \texttt{samuel.sohn@rutgers.edu} \\\And
  Sten Knutsen \\
  Department of Psychology \& \\
  Center for Cognitive Science\\
  Rutgers University --\\
  New Brunswick \\
  \texttt{sten.knutsen@rutgers.edu} \\ \And
  Karin Stromswold \\
  Department of Psychology \& \\
  Center for Cognitive Science\\
  Rutgers University --\\
  New Brunswick \\
  \texttt{karin@ruccs.rutgers.edu} \\}

\begin{document}
\maketitle
\begin{abstract}

Prosody plays a crucial role in speech perception, influencing both human understanding and automatic speech recognition (ASR) systems. Despite its importance, prosodic stress remains under-studied due to the challenge of efficiently analyzing it. This study explores fine-tuning OpenAI’s Whisper large-v2 ASR model to recognize phrasal, lexical, and contrastive stress in speech. Using a dataset of 66 native English speakers, including male, female, neurotypical, and neurodivergent individuals, we assess the model’s ability to generalize stress patterns and classify speakers by neurotype and gender based on brief speech samples. Our results highlight near-human accuracy in ASR performance across all three stress types and near-perfect precision in classifying gender and neurotype.
% disparities in ASR performance across demographic groups, underscoring existing biases in foundation models and the need for more inclusive speech recognition systems.
By improving prosody-aware ASR, this work contributes to equitable and robust transcription technologies for diverse populations.
\end{abstract}

\newcommand{\replace}[2]{#2}

\section{Introduction}

The accurate transcription of speech recordings for diverse populations is essential for advancing inclusivity in communication technologies. Prosody plays a critical role in how both humans and Automatic Speech Recognition (ASR) systems process spoken sentences, influencing interpretations of meaning and syntactic structure \cite{beach1991interpretation,snedeker2003using,carlson2009prosody,ngueajio2022hey}.
% For instance, prosodic features like phrasal stress can differentiate compound words from their adjective-noun counterparts (e.g., ``greenhouse'' vs. ``green house'').
 % or guide interpretations of ambiguous sentences
Despite its importance, prosody remains under-explored due to challenges in efficiently characterizing it.

This paper highlights the use of OpenAI's Whisper large-v2 model \cite{radford2023robust}, a state-of-the-art ASR system, to bridge this gap. Whisper leverages deep learning to analyze audio waveforms and decode them into text transcriptions, but it has not been explicitly trained to recognize prosodic stress. Through fine-tuning with stress-annotated data, we explore Whisper's potential to recognize various forms of prosodic stress (i.e., phrasal, lexical, and contrastive) and investigate the acoustic and linguistic factors influencing these patterns. By applying this approach to a diverse dataset, including neurotypical and neurodivergent groups across genders, we aim to assess the model's adaptability to diverse linguistic productions and its capacity for equitable transcription outcomes.

These investigations align with broader concerns about the performance of large foundation models in handling the diversity of human speech and language. Commercial ASR systems, including Whisper, often reflect biases stemming from training data that disproportionately represent dominant languages and dialects, exacerbating the digital divide for under-represented populations. This paper underscores the need for innovative methodologies and technologies to better account for speech and language diversity, ultimately contributing to the development of more inclusive and effective speech recognition systems.
% To this end, we investigate these challenges through two tasks: (1) learning phrasal, lexical, and contrastive stress individually to evaluate how well acoustic features generalize across stress types, and (2) classifying individuals by gender and neurotype based on a 1.7-second recording of their phrasal stress.
To this end, we investigate these challenges through two tasks: (1) learning phrasal, lexical, and contrastive stress individually to evaluate how well the acoustic features generalize across stress types for neurotypical individuals, and (2) classifying individuals as either neurotypical males (NT-M), neurotypical females (NT-F), males with Autism Spectrum Disorder or ASD (ASD-M), and females with ASD (ASD-F) based on recordings of their phrasal stress.

\begin{table*}[!b]
% \vspace{-1.1em}
\centering
% \resizebox{0.48\textwidth}{!}{ % Adjust width to text width, height adjusts automatically
\begin{tabular}{@{}ccSSS@{}}
\toprule
\multirow{2}{*}{\begin{tabular}[c]{@{}c@{}}\textbf{Training}\\ \textbf{Stress}\end{tabular}} & \multirow{2}{*}{\rotatebox{90}{\textbf{Metric}}} & \multicolumn{3}{c}{\textbf{Testing} \textbf{Stress}} \\ \cmidrule(l){3-5} 
 &  & \multicolumn{1}{c}{\textbf{Phrasal} (SD)} & \multicolumn{1}{c}{\textbf{Lexical} (SD)} & \multicolumn{1}{c}{\textbf{Contra.} (SD)} \\ \midrule
Control & \%         & 70.7\%~(4.2) & 39.5\%~(3.6) & 49.7\%~(2.6) \\ \cmidrule{3-5}
Phrasal & $\Delta$\% & \cellcolor{lightgray}19.5\%~(2.9)$^\dagger$ & 9.1\%~(9.1) & \cellcolor{lightred}-7.7\%~(4.7)$^*$ \\
Lexical & $\Delta$\% & 3.9\%~(3.7) & \cellcolor{lightgray}47.1\%~(3.6)$^\dagger$ & 27.8\%~(4.8)$^\dagger$ \\
Contra. & $\Delta$\% & \cellcolor{lightred}-11.4\%~(2.8)$^\dagger$ & 32.4\%~(3.1)$^\dagger$ & \cellcolor{lightgray}38.9\%~(2.5)$^\dagger$ \\ \cmidrule(l){3-5}
All & \%         & 90.2\%~(2.5)$^\dagger$ & 86.6\%~(2.3)$^\dagger$ & 88.7\%~(4.1)$^\dagger$ \\ \midrule
Coders & \%      & 91.9\%~(1.6) & 88.8\%~(1.6) & 91.6\%~(1.5) \\
% RFCs & \%         & 87.3\%~(3.0) & 82.8\%~(2.1) & 84.0\%~(1.6) \\
RFCs & \%         & 86.4\%~(0.2) & 83.9\%~(0.3) & 83.7\%~(0.3) \\
\bottomrule
\end{tabular}
% }
% \vspace{-10pt}
\caption{Accuracy of control stress, all-stress, coders and RFCs, and residuals for single-stress. $^\dagger$$p$ < .01 $^*$$p$ < .05}
\label{tab:macro}
% \vspace{-1.5em}
\end{table*}

\section{Fine-tuning Dataset}
Our fine-tuning dataset is based on an experiment with 66 native English-speaking college students (18 NT-M, 18 NT-F, 12 ASD-M, and 18 ASD-F) from the mid-Atlantic U.S. \cite{knutsen2024}.
For phrasal stress, participants produced 16 adjective-noun and compound word minimal pairs embedded in sentences (e.g., ``The white board/whiteboard is dirty''). For lexical stress, they produced 16 words differing only in stress pattern (e.g., ``\textit{re}cord'' vs. ``re\textit{cord}''). For contrastive stress, they listened to 16 sentences in which either a color or animal did not match a picture (e.g., ``The black sheep has the ball'' with an image of a red sheep with a ball) and corrected the error both lexically and prosodically (e.g., ``The \textit{red} sheep has the ball'').
The ground truth transcriptions were capitalized for lexical and contrastive stress to reflect the placement of stress as it would typically occur in English (Table \ref{tab:pairs}).
Three trained native English-speaking research assistants (coders), who were blind to the utterances' stress type, also hand-coded the stress of each utterance.
Each fine-tuned model presented in this work was trained with default hyperparameters for 5 epochs using 5-fold cross validation to ensure robustness \cite{de2020cross}.

\begin{table}[!h]
\resizebox{0.48\textwidth}{!}{ % Adjust width to text width, height adjusts automatically
\begin{tabular}{@{}c|l@{}}
\toprule
\rotatebox{90}{\textbf{Stress}} & \multicolumn{1}{c}{\raisebox{1.5ex}{\textbf{Minimal Pair Transcription}}} \\ \midrule
\multirow{8}{*}{\rotatebox{90}{Phrasal}} & The $<$greenhouse / green house$>$ spoils the view. \\
 & There's a $<$darkroom / dark room$>$ in this house. \\
 & The $<$whiteboard / white board$>$ needs cleaning. \\
 & That $<$hotdog / hot dog$>$ is under the table. \\
 & A $<$blackbird / black bird$>$ just flew past. \\
 & His $<$wetsuit / wet suit$>$ is on the floor. \\
 & That $<$bluebell / blue bell$>$ is pretty. \\
 & The $<$bullseye / bull's eye$>$ is red. \\
\midrule
\multirow{8}{*}{\rotatebox{90}{Lexical}} & $<$DIFfer / deFER$>$ \\
 & $<$DIScard / disCARD$>$ \\
 & $<$DIScount / disCOUNT$>$ \\
 & $<$INcrease / inCREASE$>$ \\
 & $<$INdent / inDENT$>$ \\
 & $<$INsert / inSERT$>$ \\
 & $<$INsight / inCITE$>$ \\
 & $<$INsult / inSULT$>$ \\
\midrule
\multirow{8}{*}{\rotatebox{90}{Contrastive}} & The $<$BLACK cow / black COW$>$ has the ball. \\
 & The $<$BLACK sheep / black SHEEP$>$ has the ball. \\
 & The $<$BLUE cow / blue COW$>$ has the ball. \\
 & The $<$BLUE sheep / blue SHEEP$>$ has the ball. \\
 & The $<$RED cow / red COW$>$ has the ball. \\
 & The $<$RED sheep / red SHEEP$>$ has the ball. \\
 & The $<$WHITE cow / white COW$>$ has the ball. \\
 & The $<$WHITE sheep / white SHEEP$>$ has the ball. \\
\bottomrule
\end{tabular}
}
% \vspace{-5pt}
\caption{A list of minimal pairs by stress type.
}
\label{tab:pairs}
% \vspace{-10pt}
\end{table}

\newcommand{\pad}{}
% \begin{table}
\begin{table*}[!t]
% \vspace{-0.6em}
\centering
% \resizebox{0.65\textwidth}{!}{ % Adjust width to text width, height adjusts automatically
\begin{tabular}{@{}cSSSSS@{}}
\toprule
\multirow{2}{*}{\begin{tabular}[c]{@{}c@{}}\textbf{GT}\\ \textbf{Label}\end{tabular}} & \multicolumn{5}{c}{\textbf{Predicted Label}} \\
 % & \multicolumn{1}{c}{NT-M} & \multicolumn{1}{c}{NT-F} &\multicolumn{1}{c}{ND-M} & \multicolumn{1}{c}{ND-F} & \multicolumn{1}{c}{N/A} \\ \cmidrule(l){2-6}
 & \multicolumn{1}{c}{NT-M (SD)} & \multicolumn{1}{c}{}{NT-F (SD)} &\multicolumn{1}{c}{ASD-M (SD)} & \multicolumn{1}{c}{ASD-F (SD)} & \multicolumn{1}{c}{~~UK (SD)~~} \\ \cmidrule(l){2-6}
NT-M & \cellcolor{lightgray}64.2\%~(12.4) & 0.0\%~(0.0) & 0.0\%~(0.0) & 0.0\%~(0.0) & 35.8\%~(12.4) \\
NT-F & 0.0\%~(0.0) & \cellcolor{lightgray}65.5\%~(11.4) & 0.0\%~(0.0) & 0.4\%~(0.8) & 34.1\%~(11.4) \\
ASD-M & 8.5\%~(7.9) & 0.0\%~(0.0) & \cellcolor{lightred}26.0\%~(8.4) & 0.0\%~(0.0) & 65.4\%~(7.7) \\
ASD-F & 0.0\%~(0.0) & 1.6\%~(2.4) & 0.0\%~(0.0) & \cellcolor{lightgray}55.1\%~(11.6) & 43.3\%~(11.2) \\ \bottomrule
% NT-M & 64.2\%~(12.4) & 0.0\%~(0.0) & 0.0\%~(0.0) & 0.0\%~(0.0) & 35.8\%~(12.4) \\
% NT-F & 0.0\%~(0.0) & 65.5\%~(11.4) & 0.0\%~(0.0) & 0.4\%~(0.8) & 34.0\%~(11.4) \\
% ND-M & 8.5\%~(7.9) & 0.0\%~(0.0) & 26.0\%~(8.4) & 0.0\%~(0.0) & 65.4\%~(7.7) \\
% ND-F & 0.0\%~(0.0) & 1.6\%~(2.4) & 0.0\%~(0.0) & 55.1\%~(11.6) & 43.3\%~(11.2) \\ \bottomrule
\end{tabular}
% }
% \vspace{-10pt}
\caption{This table reports the classification accuracy of Whisper-C after 5-fold cross-validation. Its class-wise precision is near-perfect and recall is moderate for all classes except ASD-M, which has 33.3\% less data.}
% $^\dagger$$p$ < .01 $^*$$p$ < .05
\label{tab:class}
% \vspace{-1.1em}
\end{table*}
% \end{table}

\section{Acoustic Feature Transfer}
In order to assess the transfer of acoustic patterns between different types of stress, we first fine-tune a Control model using all types of stress from a single random participant (Table~\ref{tab:macro}). This equips Whisper with the minimum knowledge needed to learn the transcriptions in our fine-tuning dataset.
(Phrase stress accuracy is higher for the Control because the prosodic difference between adjective-noun vs. compound word is implicitly included in the orthography of Whisper's pre-training lexicon, e.g., ``white board'' vs. ``whiteboard''.)
For phrasal stress, we fine-tune Whisper on the control data and the phrasal stress of the training subset, producing the Phrasal model that is then tested on all types of stress in the \textit{testing} subset. This is repeated for lexical stress, contrastive stress, and the combination of all three.
Table~\ref{tab:macro} shows that all non-phrasal results for single-stress training (except phrasal$\rightarrow$ phrasal) have a statistically significant improvement in accuracy over the Control model.
Phrasal and contrastive stress models learn slightly conflicting acoustic patterns in isolation, worsening their transfer accuracy significantly, but in the all-stress model, new non-conflicting patterns are learned.
When fine-tuning on all stress, we achieve near-human accuracy w.r.t. the coders, and higher average accuracy across phrasal, lexical, and contrastive stress compared to single-stress Random Forest Classifiers (RFCs) \cite{knutsen2024}.
% The RFC is an ensemble of decision trees used to reach a single, more accurate detection of stress.

\section{Gender and Neurotype Classification}
Although Whisper was intended for transcription, we demonstrate that it (Whisper-C) can be fine-tuned on the dataset to \textit{classify} recordings into 5 classes: NT-M, NT-F, ASD-M, ASD-F, and an ``unknown'' class (UK) for ambiguous recordings.
% Whisper-C was trained 5 separate times for 5-fold cross-validation, each time using default hyperparameters and a different 20\% partition of the dataset for testing \cite{de2020cross}.
%  to ensure robustness
% Cross-validation is preferred in machine learning over a paired t-test with random data splits because it provides a more robust estimate of model performance by reporting the average accuracy over its splits, addressing (but not completely eliminating) the t-test's shortcoming of assuming independent, normally distributed errors, which can lead to more inflated Type I error rates \cite{arlot2010survey,dietterich1998approximate,de2020cross}.
Table \ref{tab:class} shows that Whisper-C achieved near-perfect precision and moderate recall across all known categories except ASD-M.
Using just one recording with a mean duration of 1.7 (SD=0.4) seconds, 55.4\% of all cases were correctly classified.
Ambiguous inputs were effectively categorized as UK, with this fallback mechanism capturing 42.6\% of all cases and safeguarding the reliability of the classification pipeline.
We aggregated Table \ref{tab:class} into separate 2$\times$2 confusion matrices for M-F and NT-ASD by considering all UK cases as misclassifications (for a worst-case analysis).
Fisher's exact test yielded $p$-values of < 0.00001 for M-F and < 0.0042 for NT-ASD, indicating highly significant results.
We partly attribute the low precision and recall for ASD-M to there being 33.3\% less data than other classes.
This coincidental class imbalance is dramatically amplified in large-scale datasets used to train ASR models \cite{ngueajio2022hey}.

\section{Discussion}
This study highlights the potential of large speech models to adapt to diverse linguistic phenomena while advancing our understanding of human communication. Whisper's near-human accuracy in identifying stress patterns during transcription demonstrates its feasibility for integrating diverse prosodic features into large-scale models. The observed bidirectional transfer effects between lexical and contrastive stress patterns reveal shared acoustic underpinnings, while the weak transfer for phrasal stress (which relies predominantly on duration \cite{knutsen2024}) raises \replace{questions about handling nuanced prosodic features}{the importance of training on multiple types of stress simultaneously}.

The findings bridge psycholinguistic research and societal impact by addressing inclusivity and representation in speech technologies. Tailored transcription models for specific user groups, such as neurodivergent individuals, have potential to address the digital divide and promote equitable access. Whisper's ability to analyze prosody informs theoretical frameworks, such as the integration of prosody with syntax and semantics, and supports transcription of lesser-studied languages and dialects \cite{ladd2008intonational}.

In summary, this work underscores the importance of developing inclusive and representative speech and language models by addressing user diversity, language adaptation, and societal disparities in technology access.

% \section*{Acknowledgements}

\bibliography{main}

\begin{thebibliography}{8}
\providecommand{\natexlab}[1]{#1}

\bibitem[{Beach(1991)}]{beach1991interpretation}
C.~M. Beach. 1991.
\newblock The interpretation of prosodic patterns at points of syntactic structure ambiguity: Evidence for cue trading relations.
\newblock In \emph{Journal of memory and language}, pages 644--663.

\bibitem[{Carlson(2009)}]{carlson2009prosody}
K.~Carlson. 2009.
\newblock How prosody influences sentence comprehension.
\newblock In \emph{Language and Linguistics Compass}.

\bibitem[{De~Rooij and Weeda(2020)}]{de2020cross}
M.~De~Rooij and W.~Weeda. 2020.
\newblock Cross-validation: A method every psychologist should know.
\newblock In \emph{Advances in Methods and Practices in Psychological Science}.

\bibitem[{Knutsen and Stromswold(2024)}]{knutsen2024}
S.~Knutsen and K.~Stromswold. 2024.
\newblock Gender differences in the acoustic realization of stress.
\newblock In \emph{Penn Working Papers in Linguistics}.

\bibitem[{Ladd(2008)}]{ladd2008intonational}
D.~R. Ladd. 2008.
\newblock Intonational phonology.
\newblock In \emph{Cambridge University Press}.

\bibitem[{Ngueajio and Washington(2022)}]{ngueajio2022hey}
M.~K. Ngueajio and G.~Washington. 2022.
\newblock Hey asr system! why aren’t you more inclusive?
\newblock In \emph{International Conference on Human-Computer Interaction}.

\bibitem[{Radford et~al.(2023)Radford, Kim, Xu, Brockman, McLeavey, and Sutskever}]{radford2023robust}
A.~Radford, J.~W. Kim, T.~Xu, G.~Brockman, C.~McLeavey, and I.~Sutskever. 2023.
\newblock Robust speech recognition via large-scale weak supervision.
\newblock In \emph{International Conference on Machine Learning}.

\bibitem[{Snedeker and Trueswell(2003)}]{snedeker2003using}
J.~Snedeker and J.~Trueswell. 2003.
\newblock Using prosody to avoid ambiguity: Effects of speaker awareness and referential context.
\newblock In \emph{Journal of Memory and language}.

\end{thebibliography}

% \appendix

% \section{Example Appendix}
% \label{sec:appendix}

% This is an appendix.

\end{document}